\documentclass[12pt]{article}
\usepackage{amssymb,amsmath,graphicx}
\usepackage[mathscr]{euscript}

\def\H{{\phi}}

\def\V{{\mathscr V}}

\def\g{{ G}}
\def\h{{ h}}

\begin{document}
\begin{titlepage}
\begin{center}

{\Large A bulk inflaton from large-volume extra dimensions}

\vspace{6mm}

\renewcommand\thefootnote{\mbox{$\fnsymbol{footnote}$}}
Brian Greene${}^{1,2}$\footnote{greene@physics.columbia.edu},
Daniel Kabat${}^{3}$\footnote{daniel.kabat@lehman.cuny.edu},
Janna Levin${}^{1,4}$\footnote{janna@astro.columbia.edu}
and Dylan Thurston${}^{5}$\footnote{dpt@cpw.math.columbia.edu}

\vspace{4mm}

${}^1${\small \sl Institute for Strings, Cosmology and Astroparticle Physics (ISCAP)} \\
{\small \sl Columbia University, New York NY 10027 USA}

${}^2${\small \sl Departments of Physics and Mathematics} \\
{\small \sl Columbia University, New York NY 10027 USA}

${}^3${\small \sl Department of Physics and Astronomy} \\
{\small \sl Lehman College, CUNY, Bronx NY 10468 USA}

${}^4${\small \sl Department of Physics and Astronomy} \\
{\small \sl Barnard College, Columbia University, New York NY 10027 USA}

${}^5${\small \sl Department of Mathematics} \\
{\small \sl Barnard College, Columbia University, New York NY 10027 USA}

\end{center}

\vspace{8mm}

\noindent
The universe may have extra spatial dimensions with large volume that
we cannot perceive because the energy required to excite
modes in the extra directions is too high.  Many examples are known of manifolds
with a large volume and a large mass gap.  These compactifications can
help explain the weakness of four-dimensional gravity and, as we show here,
they also have the capacity to produce reasonable potentials for an inflaton field.
Modeling the inflaton as a bulk scalar field, it becomes very weakly coupled in
four dimensions and this enables us to build
phenomenologically acceptable inflationary models with tunings at the
few per mil level.  We speculate on dark matter candidates and the
possibility of braneless models in this setting.

\end{titlepage}
\setcounter{footnote}{0}
\renewcommand\thefootnote{\mbox{\arabic{footnote}}}

\section{Introduction}

Modern theories suggest that although the universe 
appears to have three spatial dimensions, there may in fact be more.
As is well-known, if the extra dimensions are sufficiently small, they would escape observation. If the extra dimensional volume
were large, however, a number of attractive features emerge, including an appealing explanation for the small value of Newton's constant. 
But familiar intuition suggests that as the
internal volume
grows, it becomes energetically easier to excite modes in the
extra directions--the mass gap to the Kaluza-Klein states
decreases.  This raises the question: why does the universe appear to be
three dimensional? Or, put another way, why haven't we
seen the Kaluza-Klein states?

A standard response to this question is to focus on fields that are
localized on a 3-brane so they do not probe the Kaluza-Klein states.
However, as an alternative response, we point to an
infinite number of examples that circumvent
the familiar intuition.  We will discuss known examples of
spaces that have a large mass gap
{\it and} a large volume. Consequently, even fields that did live in
the bulk would find the lowest Kaluza-Klein state energetically difficult to excite.

\begin{figure}
\centering
\includegraphics[width=90mm]{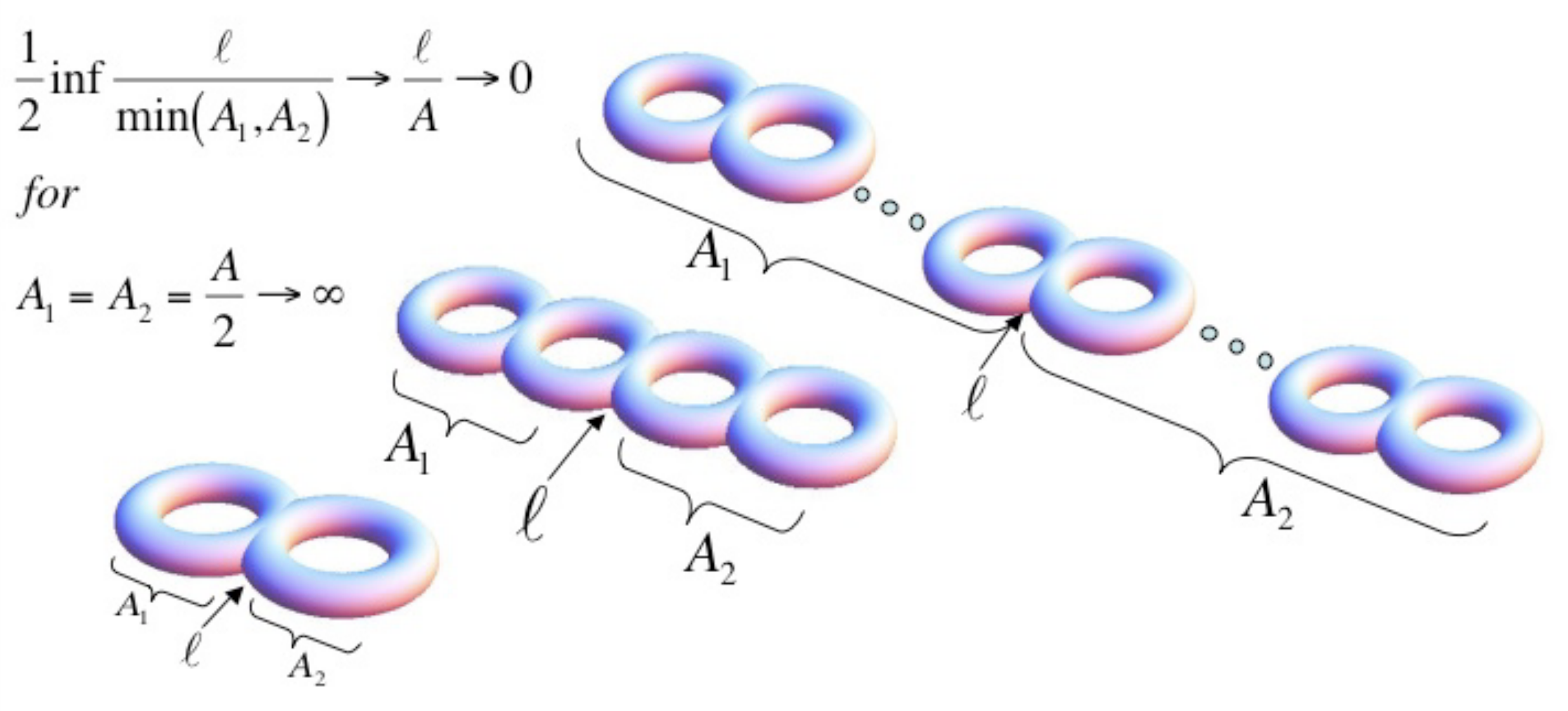}
\caption{The lowest mode on the surface made by linking doughnuts
together wobbles between the two halves divided by the curve
$\ell$.  As links are added, the tone gets lower as the two
symmetric areas grow.
\label{doughnuts}}
\end{figure}

The essential reason why some surfaces have large minimum
eigenvalue is related to the question famously posed by Mark Kac
in the 1966 paper, ``Can you hear the shape of a drum''
\cite{Kac}. While two drums can sound the same, as was shown nearly 30
years later \cite{CarolynGordon}, some features of the drum can be
heard--you can ring out the
eigenmodes of the Laplacian by banging the manifold
\cite{cornish-1999,Cornish:1997an}.  A reasonable
guess is that the bigger the drum the lower the
tone. For instance, imagine the lowest frequencies on a surface made
from stringing together doughnuts as drawn in Fig.\
\ref{doughnuts}.
The lowest tone will result when roughly half of the
surface wobbles out of phase with the other half. This conforms with
Cheeger's bound on the minimum eigenvalue \cite{Cheeger}, which for a
two-dimensional surface has the form,
\begin{equation}
k_{1}\ge \frac{1}{2}\inf \frac{{\ell}}{{\rm min}(A_1,A_2)}
\end{equation}
where $\ell$ is the length of a path that divides the surface into two
areas, $A_1$ and $A_2$, and the infimum is taken over all area
dividing paths.  In the case of the string of doughnuts, the minimum
(non-zero) eigenvalue does indeed go down with area. The larger the
area, the lower the tone.

However, there are counter-examples. For instance there are hyperbolic spaces, as
we'll elaborate, that correspond to large mass gap and large
volume.  Compactification on these spaces, and the associated cosmology, has been studied in \cite{Kaloper:2000jb,Starkman:2000dy,Starkman:2001xu,Nasri:2002rx}.  While in two dimensions these spaces are topologically equivalent to a string of
doughnuts, they are not metrically equivalent. There are no thin
bottlenecks that divide the space into roughly equal parts, so there
is no mode that wobbles a large area of the surface at once. The
lowest tone amounts to wobbling a small area. In another analogy, like
waves in a pond full of barriers, the eigenmodes can only excite small
areas at a time due to the intricate arrangement of holes.  No matter
how big you make the drum by adding more handles and holes, the lowest
tone does not get any lower.

Large-volume extra dimensions can be put to good use in diluting the
strength of gravity, thereby accounting for the small value of
Newton's constant.  Besides this phenomenological advantage, they are
a curious intellectual possibility: at every point in space there
might be some large transverse volume that we simply cannot perceive,
not because we're confined to a brane, and not because the internal
dimensions are small, but because it is simply too costly to do so at
the low energies of our everyday experience.  We discuss the
mathematical constructions in \S \ref{lv}.

As additional motivation for considering these spaces, they provide an
attractive inflaton in the form of a bulk
scalar field.  We discuss this in general in \S \ref{bulkscalar} and
study a concrete model in \S \ref{slowroll}.  Inflation in a large
volume, large gap compactification has the following attractive
features: (1) a suppression of the 4d coupling constant so the
inflaton potential is flattened, (2) a 4d description which remains
valid, even during inflation, thanks to the large gap, (3) a 4d vacuum
expectation value (vev) for the inflaton driven up to the 4d Planck
scale $M_4$, (4) an inflaton mass at the fundamental scale of the
bulk $M$, (5) inflation which takes place at an intermediate energy
density $\sim M^2 M_4^2$, and (6) a standard cosmological evolution
protected from copious and disruptive KK mode production by 
energetics.

These models have some more speculative advantages.  The inflaton is
very weakly coupled, which means it can double as a dark matter
candidate.  It is also tempting to revive the Kaluza-Klein idea in
this context and construct a braneless model in which we are
prohibited from detecting the extra dimensions by the large mass gap.
We return to these possibilities in \S \ref{summary}.

\section{Large volume, large mass gap}
\label{lv}

First we review the familiar arguments about the energetic expense of
exciting modes in the internal space.  Consider the action for a scalar
field in higher dimensions.
\begin{equation}
\int d^{N+1}x\sqrt{-\g}M^n \left (-{1 \over 2} \g^{IJ}\partial_I \H \partial_J\H
- {1 \over 2} m^2 \H^2\right )
\label{free}
\end{equation}
Here $M$ is the fundamental scale of the higher-dimensional
description; we have included an overall factor of $M^n$ so that
all fields, masses, and coupling constants will have the same units as
in $(3+1)$-dimensions.  Take a product geometry for the
$N=3+n$ spatial dimensions ${\mathbb R}^3 \times {\cal M}^{(n)}$, with metric
\begin{equation}
ds^2=G_{IJ}dx^Idx^J= \eta_{\mu \nu} dx^\mu dx^\nu + b^2 \h_{ij} dy^i dy^j
\label{metric}
\end{equation}
Here $\mu,\nu=0 \ldots 3$ and $i,j=4 \ldots N$, and
we have pulled out of the internal metric a dimensionful scale
factor $b$.
As usual this leads to a
Kaluza-Klein tower of massive states, $m_k^2=m^2+k^2/b^{2}$, where $k^2$ is a dimensionless eigenvalue
of the Laplacian on ${\cal M}^{(n)}$.  For instance, in the case of
a circle $S^1$ of size $b$, the masses
are $m \sim n/b$ for $n\in {\mathbb Z}$, which illustrates the well-known fact that for larger $b$ the modes are easier to excite.
In the absence of a brane,
the circle would have to be smaller than $b \sim {\rm TeV}^{-1} \sim
10^{-16} \, {\rm mm}$ to hide excitations of standard model fields from
experiments.

There is, however, an alternative mechanism for hiding the Kaluza-Klein modes
\cite{Kaloper:2000jb,Starkman:2000dy,Starkman:2001xu,Nasri:2002rx}.
The intuition that the minimum energy mode will necessarily decrease into an observable domain
as the volume of the internal space increases cannot
be applied to all manifolds. Indeed there are an infinite number of
manifolds whose minimum eigenvalue is {\it large}, implying a
large mass gap, despite a having large volume. We consider these now.

First consider hyperbolic
space~${\cal H}^n$ (with curvature~$-1$).  In
$n$ dimensions the square-integrable eigenvalue spectrum of ${\cal
H}^n$ is $k\in[(n-1)/2,\infty]$.  The corresponding eigenmodes define a
complete set of states in which to expand the function $\H$. Although
these square-integrable eigenmodes do vary over lengths greater than
the curvature radius, correlations beyond the curvature radius are
exponentially damped.  For this reason, these square-integrable modes
are often referred to as sub-curvature modes.

There are also super-curvature modes, modes with eigenvalues
$k<(n-1)/2$.  These correspond to eigenmodes that are not
square-integrable on ${\cal H}^n$ and are generally not considered in
the expansion of fields.  So it might seem as though there is an
intrinsic mass gap even for the simply connected infinite hyperbolic
plane: could we live with a transverse ${\cal H}^n$ and not know it?
But as mathematicians and physicists have both emphasized (see
\cite{Lyth:1995cw} and references therein), physical processes that
generate random Gaussian fields in the early universe require
contributions from both sub-curvature and super-curvature modes. We
might therefore expect cosmological processes to probe the light part
of the spectrum down to $k=0$ for the infinite hyperbolic spaces, in
which case we could not hide from the existence of the extra
dimensions.

In order to hide the extra dimensions we now consider compact hyperbolic
surfaces ($n=2$).  That is, we consider two-dimensional surfaces
with constant negative curvature as summarized in the Ricci scalar
${\cal R} = - 2/b^2$.  The Gauss-Bonnet theorem connects the area of
these spaces with their topology,
$
A=4\pi\left (g-1\right )b^2
$
where $g$ is the genus and $b$, again, is a dimensionful scale factor.
The larger the genus, the larger the area of the surface for the same
value of $b$. In most familiar examples, such as the string of
doughnuts, the minimum eigenvalue goes down with the area for fixed
$b$.  But there is an extensive literature on the construction of
hyperbolic surfaces
of arbitrary genus that possess a large
first eigenvalue: large in the sense that the lowest non-zero
eigenvalue is bounded below by the curvature scale $b^{-2}$, and is
independent of the area even as the area goes to infinity for fixed $b$
\cite{BrooksMakover1,BrooksMakover2,Buser1,Buser2,Selberg,Sarnak}.

In studying these surfaces it was originally conjectured by Buser in 1978
that the minimum eigenvalue~$k_1$ would go to zero for large genus
\cite{Buser1}. However, he later disproved his own conjecture by
exhibiting surfaces of arbitrarily large genus with minimum
eigenvalue
squared $k^2_1 \ge 3/16$ \cite{Buser2}.  The surfaces in Buser's proof
come from number theoretic
constructions. This therefore gives us hyperbolic
surfaces with arbitrarily large genus $g$, and correspondingly
large area, that maintain a large mass gap, to use the physics
lexicon.
Since the work of Buser, the number-theoretic lower bound has been
improved slightly to $k_1^2 \ge 171/784$ (for the same surfaces)
\cite{LRS}, while the construction was improved by Brooks and
Makover to allow surfaces of arbitrary genus with first eigenvalue
obeying nearly the same bound \cite{BrooksMakover1}.
If Selberg's conjecture that the square of the
minimum can be replaced by 1/4 
\cite{Selberg} is ever proven, then the theorem of Refs.\
\cite{BrooksMakover1,BrooksMakover2} would deliver the bound
$
k_{1}^2\ge 1/4
$
for these same surfaces.

In a separate construction, Brooks and Makover show that in fact a
\emph{random} surface has large first eigenvalue.  More precisely,
take a large number~$N$ of equilateral triangles and glue them
together in a random way by pairing up the edges to obtain a
triangulated surface.  The resulting surface has a canonical conformal
structure, and by the Uniformization Theorem there is a unique
hyperbolic metric in the conformal class.  Then there is a constant
$C$ so that this hyperbolic metric will satisfy $k_1^2 \ge C$ with a
probability that goes to~$1$ as $N$ goes to infinity.  (However, they
do not give an explicit value for~$C$, and their proof would probably
give a very bad bound.)  This shows that for surfaces that are
``random'' in a certain sense the first eigenvalue behaves moderately
well.

For our purposes, it is more important that we have a good bound on
$k_1^2$ than that the surfaces be generic.  We therefore continue with
the number-theoretic surfaces, and will use Selberg's
conjectured bound $k_1^2 \ge 1/4$,  although the difference between $171/784$
and $1/4$ is negligible for our purposes.

In practice then, there are surfaces of arbitrarily large genus, with
area $A\sim 4\pi gb^2$ and a minimum eigenvalue bounded from below.  For $b={\rm TeV}^{-1}$ the mass gap,
$kb^{-1} \sim {\rm TeV}$ is too large to overcome except in the
highest energy settings and yet the area is large if $g$ is large. For
$g\sim 10^{30}$, $A\sim ({\rm mm})^2 $. Despite such a large area,
we would be unable to excite modes in the higher dimensions and would
experience a 4d universe. Only at the energy scales of the Large
Hadron Collider (LHC) could we expect to witness excitation of modes
in the bulk.

These 2-surfaces are illustrative but there are presumably similar
constructions in higher dimensions.  Three
dimensional hyperbolic internal spaces of arbitrarily large volume are
known \cite{WThurston} and have the particularly nice feature of being
rigid -- all metrical quantities are fixed by the topology and the
requirement of constant curvature
\cite{mostow}.  In other words, if the volume is stabilized, all
moduli would be stabilized as a result of the rigidity.

So far, we have consider only the Laplacian (scalar) spectrum.
Spinors also need to see a large mass gap in a realistic theory. The
Dirac eigenspectrum is less well studied and it is not yet known if the
large genus hyperbolic surfaces discussed above have a suitable
spectrum.  Ammann, Humbert, and Jammes have constructed surfaces (of
any genus, with bounded volume) with a zero mode followed by an
arbitrarily large gap in the Dirac spectrum
\cite{ammann}, although these
surfaces (dubbed ``Pinocchio surfaces'', formed by stretching out a
long nose from the surface) do not have a suitable Laplacian spectrum.

Although we have focused on hyperbolic spaces, there are other constructions.
For instance one can obtain a large gap on a flat 2-torus, simply by allowing the complex structure to degenerate \cite{Dienes}.  Another example, which gives the desired Kaluza-Klein
tower for both scalars and fermions, is a rectangular $n$-torus of
volume $\sim b^n$ with $n\gg 1$. The mass gap stays fixed even as the
volume can be sent to infinity by sending the number of dimensions to
infinity.  This is less remarkable than the hyperbolic
construction: each individual direction is small and the large volume
is simply a result of a large number of dimensions.  Also
there is a huge spinor degeneracy since the
number of spinors grows exponentially with the number of dimensions.
Still, the $n$-dimensional torus
demonstrates the existence of a space that has the required large mass
gap for both scalars and fermions.

\section{Bulk Inflation}
\label{bulkscalar}

One phenomenological advantage to having a large volume is that it
weakens the observed force of gravity in four dimensions.  But any
other bulk interactions will be suppressed as well.  In this section
we use this to help construct inflationary potentials \cite{Mazumdar:1999tk}.

We begin with a $\phi^4$ theory in the bulk, with action 
\begin{align}
&\int d^{4+n}x\sqrt{-\g} M^n\left
[-{1 \over 2}\g^{IJ}\partial_I \H_B \partial_J\H_B - {1 \over 4} \lambda_{B}\left (\H_B^2-v_B^2\right )^2\right ] 
\label{interact}
\end{align}
where bulk quantities carry a $B$.  Integrating over the internal
dimensions the action becomes
\begin{equation}
\int d^{4}x\sqrt{-g} \left
[-{1 \over 2} g^{\mu \nu}\partial_\mu \H \partial_{\nu}\H - {1 \over 4} \lambda\left (\H^2-v^2\right )^2\right ] \quad ,
\end{equation}
where we canonically normalize the kinetic term by redefining $\H =
\V^{1/2} \H_B$.  Here
\begin{equation}
\V=b^n M^n \int d^n y\sqrt{h}
\end{equation}
is a dimensionless measure of the volume of the internal space, and
the 4d coupling and vev are related to the bulk values through
\begin{align}
\label{lam}\lambda &=\lambda_B \V^{-1}\\
\label{vee}v^2 &=v_B^2 \V\quad .
\end{align}
It follows that the mass is the same in the bulk and 4d descriptions.
\begin{equation}
m^2=m_B^2=\lambda_B v_B^2 \quad.
\label{m2}
\end{equation}
These simple equations highlight the main features of large-volume
compactification: we naturally get models with tiny couplings and huge
vevs.

To get a sense of scale we compare to the gravitational action under dimensional reduction.
\begin{equation}
\label{HigherDim}
\int d^{4+n}x\sqrt{-G} \, \frac{1}{2}M^{2+n}{\cal R} \quad \rightarrow \quad
\int d^4x\sqrt{-g}\left[\frac{1}{2}M_{4}^2{\cal R}^{(4)}+...\right ]
\end{equation}
Here $M$ is the underlying higher-dimensional scale and the effective
reduced four-dimensional Planck mass is
\begin{equation}
M^2_{4}=M^{2} \V \sim \left(10^{18} \, {\rm GeV}\right)^2 \,.
\label{planck}
\end{equation}
Leaving $M$ the unknown, this requires the volume adjust by
$\V=M^2_{4}/M^2$.  If the bulk coupling constant $\lambda_B \sim {\cal
O}(1)$ and the bulk vev $v_B = {\cal O}(M)$ then
\begin{align}
\lambda &\sim \left(M/M_4\right)^2
\nonumber \\
v &\sim M_{4}
\nonumber \\
m &\sim M \quad.
\label{setscale}
\end{align}
Taking $M \sim {\rm TeV}$, for example,
the coupling in the 4d theory is minute.  The vev is at the 4d
Planck scale, while the mass is much below Planck scale.
Intriguingly, this implies that if there exist fundamental scalar
fields in the bulk their interactions should be brutally suppressed.
We would not easily observe such scalar fields, as indeed we do
not. Furthermore, any scalar field potential would be exceedingly flat
as slow-roll inflation requires: a very small coupling and a very
large vacuum expectation value. And, neatly enough, any remnant scalar
particles from the early universe would be dark matter candidates,
with a mass set by the underlying higher-dimensional Planck scale $M$.

We note that although $\H$ has mass set by the bulk scale $M$,
inflation occurs at a much higher energy scale. Near the maximum of
the potential, where $\H\ll v$, the effective 4d energy density is
\begin{equation}
V=\lambda v^4= \V \lambda_B v_B^4 \sim M^2 M_4^2 \sim \left (10^{10} \text{GeV} \right )^4
\label{scale}
\end{equation}
where we're assuming the bulk energy density $\lambda_B v_B^4 \sim M^4
\sim {\rm TeV}^4$.  So an intriguing observation about the inflaton
potential is that the energy scale of inflation would be $10^{10} \,
{\rm GeV}$ despite being driven by a field with an electroweak scale
mass.

Although the choice $M \sim {\rm TeV}$ is natural from the point of
view of electroweak physics, the resulting inflationary scale $\sim
10^{10} \, {\rm GeV}$ does not generate density perturbations of the
required magnitude.  Instead, as we'll see in the next section, the
observed density perturbations favor the existence of an intermediate
fundamental scale, with $M \sim 10^{11} \, {\rm GeV}$ and ${\cal V}
\sim 10^{14}$.

\section{A slow-roll model}
\label{slowroll}

In this section we study a concrete model of bulk inflation and show
that we can get a reasonable power spectrum, density perturbations of
the right magnitude, and the requisite number of $e$-folds, all with
tunings of the inflaton potential at the few per mil level.

\begin{figure}
\centering
\includegraphics[width=70mm]{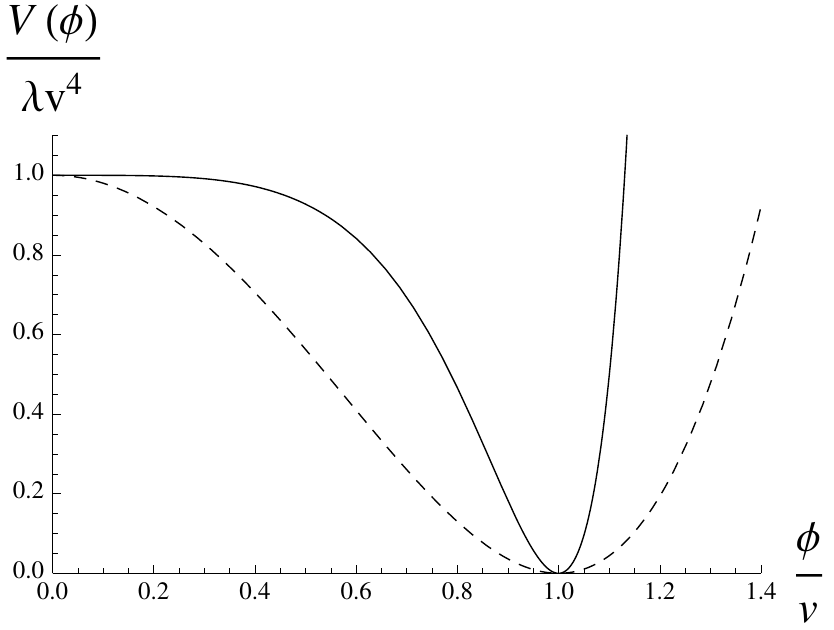}
\hfill
\caption{Top dotted line has $\alpha=0$ and so is a simple $\H^4$ style
potential.  The solid line has $\alpha=1$.\label{potentials}}
\end{figure}

We emphasize that any reasonable potential could be chosen for the
inflaton.  For simplicity we take a potential of the form
\begin{equation}
V={1 \over 4}\lambda e^{2\alpha \H^2/v^2}\left (\H^2-v^2\right )^2
\label{given}
\end{equation}
where $\lambda,v$ are set as in (\ref{lam}), (\ref{vee}).  Setting
$\alpha=0$ recovers the usual $\phi^4$ potential, while setting $\alpha = 1$ makes the second derivative of the
potential vanish at the origin (see Fig.\ \ref{potentials}).  We could equally
well have used a potential of the Coleman-Weinberg type
\cite{1973PhRvD...7.1888C} or any other variant of inflaton potential.
There are various phenomenological constraints that must be satisfied.

\medskip\noindent {\em Power spectrum} \\
First we quantify the naturalness of $V$ as a slow-roll inflaton
potential using the parameters described in \cite{slow}.
Slow-roll inflation is a consistent assumption if the slope and the
curvature of the potential are small as quantified by the slow-roll
parameters $\epsilon$ and $\eta$.  Denoting $v = \beta M_4$, for small-field inflation $\H \ll v$
and we have
\begin{align}
\nonumber
&\epsilon = \frac{M_{4}^2}{2}\left (\frac{V^\prime}{V}\right )^2 \approx \frac{8}{\beta^2}\frac{\H^2}{v^2}\left[
\alpha-1-\frac{\H^2}{v^2}\right ]^2 \\
&\eta - \epsilon = M_4^2 \left(\frac{V'}{V}\right)'  \approx -\frac{4}{\beta^2}\left[1-\alpha+3\frac{\H^2}{v^2} \right ]\,.
\end{align}
We now study two special cases in turn.  



$\bullet $ $\alpha < 1$:

When $\H\ll v$ we have
\begin{align}
\epsilon &\ll |\eta|
\nonumber \\
\eta &\sim-\frac{4}{\beta^2}(1 - \alpha)
\end{align}
leading to a power spectrum $P_S\propto k^{n_S-1}$ with scalar spectral index
\begin{align}
n_S = 1 - 4\epsilon + 2\eta \approx 1 -\frac{8}{\beta^2}(1 - \alpha) \,.
\end{align}
So for $\alpha < 1$ we have a red spectrum.  In fact for generic
values of $\alpha < 1$ the spectrum of scalar fluctuations is too red
unless $v \gg M_{4}$ -- the usual issue for quartic potentials for
massive inflatons.  Requiring that $n_S>0.95$ for $\alpha = 0$, for
instance, would demand the uncomfortable value $\beta = v / M_4 > 12$.
As an alternative to a trans-Planckian vev one can tune $\alpha$ close
to 1.  For instance taking $\beta = 1$ and requiring $n_S > 0.95$
implies $1 - \alpha < 6 \times 10^{-3}$.

$\bullet $ $\alpha = 1$:

Setting $\alpha = 1$ and taking $\H\ll v$ we have
\begin{align}
\epsilon &\ll |\eta|
\nonumber \\
\eta &\sim -\frac{12}{\beta^2}\frac{\H^2}{v^2}
\end{align}
leading to
\begin{equation}
n_S = 1 - \frac{24}{\beta^2} \frac{\H^2}{v^2}\,.
\end{equation}
Provided inflation occurs at sufficiently small $\H$ this is an
acceptable, slightly red spectrum.  For example, as we'll see below,
taking inflation to begin at $\H = 0.04 \, M$ leads to a reasonable
number of $e$-folds.  For $\beta = 1$ this leads to $n_S = 0.96$.
But achieving this does require some fine-tuning of the potential.  For
the approximation $\alpha = 1$ to be valid we need
\begin{equation}
1 - \alpha < 3 \H^2/v^2
\end{equation}
which for the values mentioned above leads to
\begin{equation}
1 - \alpha < 5 \times 10^{-3}\,.
\end{equation}

\medskip\noindent {\em Number of e-folds} \\
The number of e-folds is given by 
\begin{equation}
N = \frac{1}{M_{4}} \int_{\H}^{\H_e}
\frac{d\H}{\sqrt{2 \epsilon(\H)}}\,.
\end{equation}
As a concrete example, consider taking $\alpha = 1$, so that
\begin{equation}
N =\frac{\beta^2}{8}\left (\frac{v^2}{\H^2}-\frac{v^2}{\H_e^2}\right )
\end{equation}
Slow roll inflation ends when $\epsilon \sim 1$, or roughly when the
field settles into its minimum, so that $\H_e \sim v$. Sufficient
e-folds then requires $\phi_i < \beta v/\sqrt{8N}$. For $N \sim 60$
and $\beta \sim 1$ the condition amounts to $\phi_i< v / 20$ which is
reasonable.

\medskip\noindent {\em Density perturbations} \\
Density perturbations are crucial in determining the energy density
during inflation.  In our case they will set the value of $M$ or
equivalently $\V$. The size of scalar perturbations is given by 
\begin{equation}
\frac{\delta \rho}{\rho} = \frac{H}{\pi M_{4}}\frac{1}{\sqrt{8 \epsilon}}
\end{equation}
During slow-roll $H^2 \approx V/3 M_{4}^2$, and
assuming inflation begins at $\H_i \ll v$, the energy density during
inflation
$V \approx {1 \over 4} \lambda v^4$.
Taking $\alpha = 1$, the slow-roll parameter $\epsilon \approx 8 M_4^2
\H^6/v^8$ so
\begin{equation}
\frac{\delta \rho}{\rho} = \frac{\lambda^{1/2} \beta^6}{16 \pi \sqrt{3}} \, \left(\frac{M_4}{\H_i}\right)^3\,.
\end{equation}
With the number of $e$-folds $N \approx \beta^4 M_4^2 / 8 \H_i^2$ we
have
\begin{equation}
\frac{\delta \rho}{\rho} = \frac{1}{\pi}\sqrt{\frac{2}{3}} \, \lambda^{1/2} N^{3/2}\,.
\end{equation}
Sufficient inflation requires $N\sim {\cal O}(60)$, and observation
requires $\delta \rho / \rho \sim 10^{-5}$.  This leads to $\lambda
\simeq 10^{-14}$.  From the four dimensional point of view this would
be viewed as a very fine-tuned coupling.  But in the context of
large-volume extra dimensions it's quite easy to achieve.  Let's take
the bulk coupling $\lambda_B = {\cal O}(1)$.  Then the required
four-dimensional coupling translates into
\begin{equation}
\V = \lambda_B/\lambda \simeq 10^{14}
\end{equation}
implying a bulk scale intermediate between the electroweak and 4d
Planck scales:
\begin{equation}
M = M_4/\sqrt{\V} \simeq 10^{11} \, {\rm GeV}\,.
\end{equation}
This implies an energy density during inflation
\begin{equation}
V = {1 \over 4} \lambda v^4 = {1 \over 4} \lambda_B \beta^4 M^2 M_4^2 \sim \left(10^{14} \, {\rm GeV}\right)^4
\end{equation}
where we've taken $\lambda_B$ and $\beta$ to be ${\cal O}(1)$.

Although this seems like the most natural way to realize bulk
inflation, there are other possibilities.  For instance we could
demand that $M \sim 1 \, {\rm TeV}$ is of order the electroweak scale.
This leads to $\V = M_4^2/M^2 \sim 10^{30}$ and, taking $\lambda_B =
{\cal O}(1)$, $\lambda = \lambda_B/\V \sim 10^{-30}$.  Then acceptable
density perturbations require an extended period of inflation, $N \sim
10^7$, which requires that inflation begin at a very small value of
$\H$: $\H_i \sim 10^{-4} \beta v$.  This can be arranged but may not
be an appealing condition.  Regardless of the particular potential or
exit method, the gist is that density perturbations in this approach
set the bulk scale, and this scale will fall somewhere between the
Planck scale and the electroweak scale depending on the details.

As another alternative to having an intermediate fundamental scale,
suppose there is only the electroweak scale and a lot of extra flat
dimensions of size $L$. Then small changes in $L$ from the time of
inflation until today would mean drastic changes in the internal
volume, allowing it to be small enough for decent density
perturbations during inflation and large enough for a heavy Planck
mass today. A $\V \sim (LM)^n\sim 10^{15}$ is accommodated easily by
$L=10^{15/n}M$, a very modest change in $L$ when $n\gg 10$. Still,
this is a form of fine tuning since if $L$ is much bigger or smaller
the density perturbations slip out of the desired range.  While we
don't defend this obvious fine-tuning, we mention that it's possible
that $L$ rolls slowly during inflation and the end of inflation
happens precisely when $L$ falls into it's potential. So the last 60
or so e-foldings, the ones we observe, happen by definition near $L$
critical. We won't pursue the details of a hybrid model here. It might
be more attractive for $M \sim {\rm TeV}$ if the perturbations could
be found naturally in a source other than the inflaton
\cite{Lyth:2002my,Dvali:2003em,ArmendarizPicon:2003ht}.

\section{Discussion and speculation}
\label{summary}

In summary, we have discussed internal manifolds with both a large
volume and a large mass gap.  From a mathematical point of view such
manifolds seem generic in the space of all compactifications.  From a
physical point of view they are interesting because the large volume
accounts for the weakness of four dimensional gravity, while the large
mass gap makes the extra dimensions invisible in current experiments.

A bulk scalar field, if present in such a compactification, has some
curious features.
In terms of the fundamental Planck scale $M$ and
dimensionless volume $\V$ of the extra dimensions we expect the 4d
field to have a mass, coupling, vev and energy density
\begin{align}
& m \sim M \\
& \lambda \sim 1/\V \nonumber \\
& v \sim \sqrt{\V} M \nonumber \\
& V \sim \V M^4 \nonumber
\end{align}
In terms of the 4d Planck mass $M_4 = \sqrt{\V} M$ this means
\begin{align}
& m \sim M_4/\sqrt{\V} \\
& \lambda \sim 1/\V \nonumber \\
& v \sim M_4 \nonumber \\
& V \sim M_4^4/\V \nonumber
\end{align}
From the 4d point of view its vev is large, of order the 4d Planck
scale: $v \sim M_4$.  But its coupling is tiny, $\lambda \sim 1/\V$, which suppresses its mass
and energy density.
These features are attractive for building inflationary potentials.
Here we comment on some of the fine-tuning issues which are involved.
As we saw in section \ref{slowroll}, it is not easy to satisfy the
slow-roll conditions and obtain an acceptable perturbation spectrum --
the so-called $\eta$ problem of inflationary cosmology.  We finessed
this by tuning the potential at the few parts per mil level.  Given
this tuning, it is fairly easy to get enough $e$-folds of inflation.
But the big payoff of a large-volume compactification is in generating
density perturbations of the right magnitude.  Normally this requires
a tiny fine-tuned coupling from the 4d point of view.  But in the
extra dimensional scenario such a coupling is quite natural, and leads
us to identify a fundamental bulk scale of perhaps $10^{11} \, {\rm
GeV}$.

Most of these features rely on having a large internal volume.  But
the large gap plays an important role as well, because we need to ask:
is the use of 4-dimensional effective field theory valid during
inflation?  In this regard it's reassuring that the energy density
during inflation $V \sim \lambda M_4^4$ is well below the 4d Planck
scale, so 4d quantum gravity effects should be
negligible. But what about the Kaluza-Klein tower?  To address this
note that for the potential (\ref{given}) the Hubble parameter during
inflation
$H^2 \approx V/3M_4^2 \approx \lambda_B \beta^4 M^2 / 12$.
This corresponds to a de Sitter temperature
\[
T = {H \over 2\pi} \approx {\sqrt{\lambda_B} \beta^2 M \over 4\pi\sqrt{3}}\,.
\]
Given the bounds discussed in \S \ref{lv} and taking $b \approx 1/M$, the
Kaluza-Klein tower begins at the scale $M/2$.  So a naive estimate is that
Kaluza-Klein excitations are suppressed by a Boltzmann factor $\exp(- 2 \pi
\sqrt{3}/\sqrt{\lambda_B} \beta^2)$.  Even for $\lambda_B \approx \beta \approx 1$
this is a suppression by almost $10^{-5}$.  By tuning $\lambda_B$ and $\beta$
to be slightly less than one -- something which is desirable in any case, to
avoid strong coupling in the bulk and a trans-Planckian vev -- the contribution of the Kaluza-Klein
tower can be made negligible.  Similar remarks apply to the effects of possible higher-derivative
terms in the bulk gravitational action (\ref{HigherDim}), which are suppressed by powers of
\[
{\cal R} / M^2 \approx 12 H^2 / M^2 \approx \lambda_B \beta^4\,.
\]
By tuning $\lambda_B$ and $\beta$ slightly these terms can be brought under control.

Putting this differently, if the mass gap were small there would be good
astrophysical reasons to be concerned that a standard cosmology would
not be possible. Although weakly coupled, Kaluza-Klein modes of the scalar field
could still be copiously produced if the volume were large and the
modes correspondingly easy to produce. Our large volume, large mass
gap manifolds provide a protective energetic barrier and allow for a
standard cosmological evolution (which resonates with the perspective of
\cite{Kaloper:2000jb}).
Thus inflation driven by a bulk scalar field seems like an attractive
possibility.  

We have implicitly assumed that the radion, and all other moduli,
are stabilized during inflation.\footnote{The radion must have a large enough
mass that its Boltzmann factor $\exp(-4 \pi \sqrt{3} M_{\rm radion} / \sqrt{\lambda_B} \beta^2 M)$
can be neglected during inflation.}  Incorporating a
mechanism for radion stabilization would be an important next step
in developing this model.  A mechanism for stabilizing moduli is required for all
higher-dimensional cosmologies, and many scenarios have been developed.
Stabilization might be achieved via twisted scalar fields \cite{Goldberger:1999uk},
string windings \cite{windings}, Casimir energy \cite{Casimir}, fluxes \cite{PhysRevD.68.046005}, or
some other motivated set of potentials
\cite{Greene:2007sa}. Also, some evolution
of the moduli could be phenomenologically interesting if both
a bulk scalar and a radion are at play in double-field
inflation. To keep our focus clear, we have not addressed
moduli stabilization, but rather defer to the long list of possible
mechanisms discussed in the literature.

We conclude with two more speculative possibilities
which may be realized within the large-volume, large-gap scenario.

First, any remnant scalar particles from the early universe would be
dark matter candidates. As a result of the suppression of the coupling
constant the particles are effectively non-interacting, that is to
say, dark.  At the end of inflation, the flow of $\H$ particles into
standard model particles through parametric resonance could
potentially overcome the very weak coupling and produce appropriate
abundances of dark and baryonic matter.

Second, and more speculatively, one could imagine constructing fully
braneless models along these lines.  That is, one could allow  all
fields -- including standard
model fields -- to propagate in the bulk.  The large volume would account
for the weakness of gravity by diluting its strength along the lines of \cite{ArkaniHamed:1998rs},
while the large gap would keep the extra dimensions from being
directly detected.  The challenges in realizing this scenario are (i)
obtaining realistic interactions since the strength of all forces
would be diluted over the large internal volume, and (ii) obtaining a realistic
spectrum of chiral fermions.  
Regarding point (i), excited
Kaluza-Klein modes are localized at around the curvature scale and so are
not diluted over the entire internal volume. Consequently their interactions can be of reasonable
strength.  One might therefore hope to model massive gauge fields
along these lines.  Regarding point (ii), we note that the
spectrum of the Dirac operator on these spaces is not well understood.
Clearly the details of the phenomenology will depend crucially on the
specific internal geometry, the eigenspectra of the various operators,
and the overlap integrals of eigenmodes.  Quantum effects
on the finite internal volume will naturally have something to say
about these issues.  In a realistic approach,
chiral fermions must be generated, the Einstein equations must be
satisfied\footnote{The curvature term from a hyperbolic internal space
would be large, $1/b^2 \approx M^2$, and would dominate the energy
density of the universe, contrary to observations.}, coupling
constants must be resuscitated, and the extra dimensions must be
stabilized.

\bigskip

\centerline{\bf Acknowledgements}

We are grateful to Gustavo Burdman, Pedro Ferreira, Kurt Hinterbichler
Eugene Lim, Maulik Parikh, Eduardo Ponton and Sarah Shandera for their
insightful comments.  DK is supported by U.S.\ National Science
Foundation grant PHY-0855582 and PSC-CUNY award \#60038-39-40.  JL
acknowledges financial support from an NSF Theoretical Physics grant,
PHY - 0758022.

\providecommand{\href}[2]{#2}\begingroup\raggedright\endgroup

\end{document}